\pdfoutput=1
\documentclass[prd, showpacs, superscriptaddress, floatfix, nofootinbib]{revtex4}

\usepackage{amsmath}
\usepackage{graphicx}

\def\be{\begin{equation}}
\def\ee{\end{equation}}
\def\bea{\begin{eqnarray}}
\def\eea{\end{eqnarray}}

\title{title: TBA}

\begin{document}

\title{Scale-Invariant Fluctuations from Galilean Genesis}

\author{Yi Wang and Robert Brandenberger}
\email{wangyi,rhb@physics.mcgill.ca}
\affiliation{Department of Physics, McGill University, Montr\'eal, QC, H3A 2T8, Canada}


\begin{abstract}

We study the spectrum of cosmological fluctuations in 
scenarios such as Galilean Genesis \cite{Nicolis} in which
a spectator scalar field acquires a scale-invariant spectrum of
perturbations during an early phase which asymptotes 
in the far past to Minkowski space-time. In the case of
minimal coupling to gravity and standard scalar field Lagrangian,
the induced curvature fluctuations depend quadratically on the
spectator field and are hence non-scale-invariant and
highly non-Gaussian. We show that if higher dimensional operators
(the same operators that lead to the $\eta$-problem for inflation)
are considered,
a linear coupling between background and spectator field
fluctuations is induced which leads to scale-invariant and
Gaussian curvature fluctuations.

\end{abstract}

\maketitle

\section{Introduction}

Inflation \cite{Guth} is currently the most well-studied and successful paradigm for 
very early universe cosmology. Most importantly, inflation is the first
scenario based on causal physics to give a mechanism \cite{Mukh} to generate 
perturbations on the scales currently observed. The predicted spectrum is
nearly scale-invariant and close to Gaussian. These predictions have been
spectacularly confirmed in recent cosmic microwave (CMB) measurements \cite{WMAP}.

As already pointed out \cite{Peebles} a decade before the development of inflationary
cosmology, any early universe scenario which yields an almost adiabatic and nearly
scale-invariant spectrum of fluctuations on scales which are super-Hubble at 
late times (e.g. the time of equal matter and radiation) will yield an angular
power spectrum of CMB anisotropies in agreement with observations. Inflation
is the first but not the only model which yields such fluctuations. 

It is valuable to investigate alternative scenarios to obtain a scale-invariant 
spectrum of curvature fluctuations using causal microphysics. On one hand, one 
does not know apriori whether inflation is indeed the right paradigm. At least 
in its current realizations based on coupling
scalar fields to Einstein gravity it suffers from a number of conceptual challenges
(see e.g. \cite{TPproblem}), and if inflation in fact turns out not to be the
correct paradigm, alternatives are certainly needed. On the other hand, even if 
inflation turns out to be the correct paradigm, competing scenarios are useful 
because they may provide guides to future experiments which will allow 
inflation to pass further non-trivial tests. For example,
the ``String Gas Cosmology" alternative to inflation \cite{SGC} predicts
a blue tilt of the spectrum of gravitational waves, whereas inflationary models
developed within the context of Einstein gravity generically produce a small red
tilt. Thus, future measurements of the slope of the spectrum of gravitational
waves could either falsify inflation or allow it to pass a further non-trivial
test. In this sense having alternatives to inflation makes early universe 
cosmology a healthier science.

Recently, there is an increasing interest in a class of alternatives to inflation 
in which the scale-invariance of the fluctuations is induced not by
the de-Sitter-like expansion of space, but by the evolution of another matter
field. Examples are the {\it conformal cosmology} of \cite{Rubakov}
(see also \cite{Rubakov2, Rubakov3, Rubakov4}), in
which the evolution of a conformal scalar field induces scale-invariant
fluctuations in an axion-like field to which it couples, 
the {\it pseudo-conformal cosmology} of \cite{Khoury} and the
{\it Galilean genesis} model of \cite{Nicolis} in which a Galileon field
which dominates the background dynamics of space-time induces scale-invariant
fluctuations of a spectator scalar fields.

In this paper we propose a mechanism by which the scale-invariance of
the spectator scalar fields can be transferred to a scale-invariant
spectrum of curvature fluctuations. The standard curvaton mechanism
does not work since in this case the curvature fluctuations are
quadratic in the spectator scalar field, and hence the fluctuations
will be neither scale-invariant nor Gaussian. The key to our
construction is to introduce a linear coupling between the nontrivial
background geometry and the spectator scalar.

\section{Emergent Dynamics from Emergent Cosmology}

The three scenarios mentioned above all can be viewed as
particular realizations of an {\it emergent cosmology}, 
in the sense that there is a spectator field whose equation
of motion looks like that of a standard scalar field in a de Sitter
background geometry, while the actual space-time is not de Sitter
at all. For example, in the {\it Galilean Genesis} scenario it is an
evolving Galileon field which couples to a spectator scalar field
like a de Sitter metric would, while the Galileon field leads to
a metric which approaches that of flat Minkowski space-time as
time tends to $t \rightarrow - \infty$. In this case, the
expanding background expansion is in fact emergent from an
initial static phase (as is also postulated to happen in String Gas
Cosmology \cite{SGC}). 

There is, however, a generic problem which arises if one
generates a scale-invariant spectrum of fluctuations of a
spectator scalar field in the hope of then obtaining
scale-invariant curvature perturbations \cite{Laurence}:

\begin{itemize}

\item If the scalar field has no potential term, the scale-invariant perturbation 
of the scalar field couples quadratically to the curvature perturbation $\zeta$
as $\dot{\zeta} \sim \delta\phi^2$, and thus the resulting curvature perturbation 
is neither scale invariant nor Gaussian. 

\item If the spectator scalar field has a potential (say, a fixed mass term) and its
background value is excited and evolving in time (like the inflation
and the curvaton are in inflationary cosmology), then - although the
background scalar field allows a linear coupling between the spectator field
fluctuations and the curvature perturbation - the background
spectator scalar field will destroy the emergent background phase in
the far past, when fluctuations are supposed to be generated.

\end{itemize}

Actually, the solution to the above problems is built-in in those theories. 
To see this, it is helpful to pause and revisit the ``$\eta$-problem'' for 
inflation \cite{Copeland:1994vg}. The $\eta$-problem states that in a 
time-dependent background, a massless field $\phi$ typically  gets a mass of 
order of the Hubble parameter, unless the masslessness is protected by 
a symmetry such as a shift symmetry \cite{Yamaguchi}. For example, in
the context of supergravity models, then if the field comes from a 
K\"ahler potential, the Lagrangian will have the structure
\begin{align}
  \mathcal{L} \supset V_0 ~\exp\left[\mathcal{O}(1) \frac{\phi^2}{M_p^2}\right]~,
\end{align}
where $V_0$ is the effective vacuum energy in the system,
and $M_p$ is the Planck mass. Expanding this equation, 
the $\phi$ field obtains a mass of order Hubble parameter $H$:
\begin{align}
  \mathcal{L}\supset \left[1 + \mathcal{O}(1) \frac{\phi^2}{M_p^2}\right] V_0 
  \simeq V_0 + \mathcal{O}(1) H^2\phi^2 ~.
\end{align}
It is believed that the above effect is so general that it becomes a problem of 
inflation. 

There is good reason to believe that the same term should also naturally 
arise in emergent universe scenarios where the Hubble parameter is emergent 
(specifically in the Galilean Genesis model). Thus, a light scalar field $\phi$ 
(our spectator scalar field) obtains 
an emergent mass. The emergent mass, on one hand, breaks the shift symmetry. 
This will induce a linear coupling of the scalar field fluctuations to curvature 
perturbations, and will allow a scale-invariant spectrum of the spectator
field to induce a scale-invariant spectrum of curvature fluctuations. On the 
other hand, the mass of the spectator field is emergent and thus the 
induced potential of the field does not destroy the initial emergent background.
Specifically, it does not break the emergent nature of expansion of the universe.

\begin{figure}[htbp]
\centering
\includegraphics[width=0.45\textwidth]{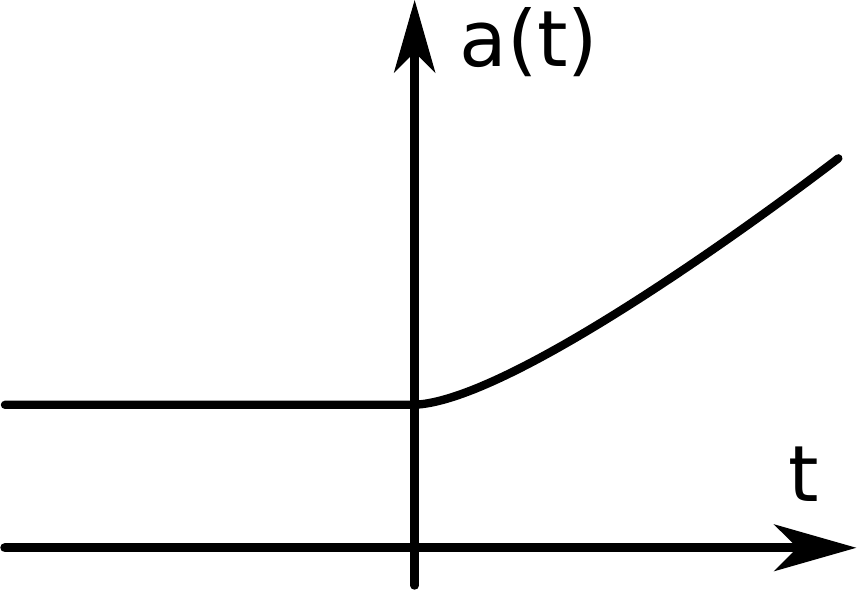}
\hspace{0.07\textwidth}
\includegraphics[width=0.45\textwidth]{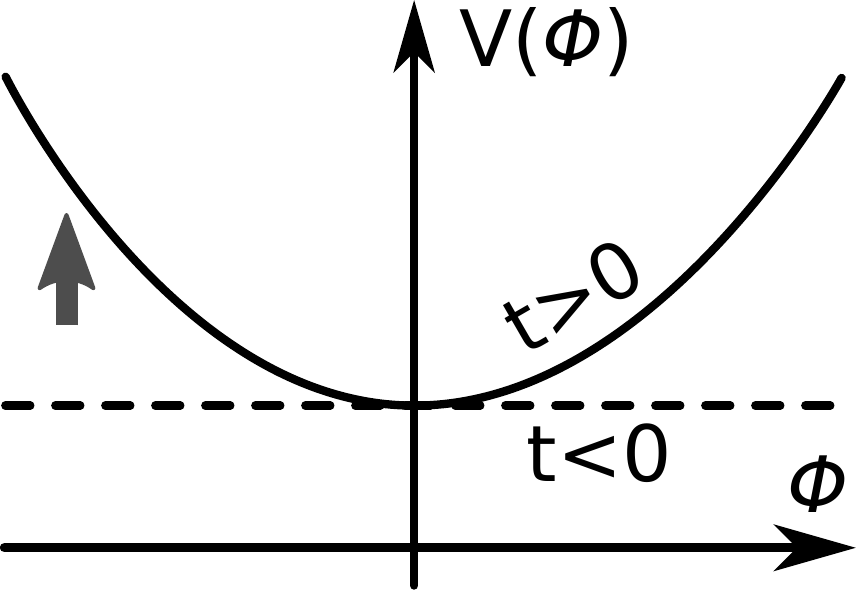}
\caption{An emergent potential from emergent cosmology.\label{fig:emergon}}
\end{figure}

For this purpose, starting with the Lagrangian $\mathcal{L}_m$ for 
a massless scalar field, we shall consider the corrected matter Lagrangian
\begin{align}
  \mathcal{L}_m \rightarrow \left(1+\frac{\phi^2}{M^2}\right)\mathcal{L}_m~,
\end{align}
where $M$ is a mass parameter characteristic of the new physics which 
yields the corrections to the Lagrangian. This mass will induce non-trivial
dynamics of $\phi$ which in turn leads to a linear coupling between 
fluctuations in $\phi$ and curvature perturbations.

Alternatively, we could also consider a gravitational action with induced
non-minimal coupling of $\phi$ to gravity (a similar mechanism was
proposed in \cite{Rubakov4}):
\begin{align}
  \frac{M_p^2}{2}\int d^4x \sqrt{-g} ~R \rightarrow 
\frac{M_p^2}{2}\int d^4x \sqrt{-g} ~R - \frac{\xi}{2} \int d^4x \sqrt{-g} ~\phi^2 R~.
\end{align}
This modified action also leads to a direct linear coupling between fluctuations
of $\phi$ and curvature perturbations.

For definiteness, in the present work, we only consider the first possibility, which is
inspired by a K\"ahler potential in supersymmetry 
\footnote{In the case of a K\"ahler potential, it is more natural to have a standard 
kinetic term, leaving the exponential 
$\exp\left[\mathcal{O}(1) \frac{\phi^2}{M_p^2}\right]$ coupling to the potential term. 
However, here we consider an over-all exponential term. For the Galilean case, we 
have to because there is no potential for the Galilean field. On the other hand, in the 
pseudo-conformal case, we could consider the case where the exponential only couples 
to the potential, and the result should not change much.}. We expect that the modified 
gravity case should have a similar effect, via a conformal transformation.

\section{Galilean genesis example}
\label{sec:galil-genes-example}

\subsection{Preliminaries}

Consider the Galilean scenario \cite{Nicolis}. In the Galilean case, the gravity sector 
is standard, with
\begin{align}
  S_g = \frac{M_p^2}{2}\int d^4x\sqrt{-g} ~R~.
\end{align}
The Galilean field $G$ has an action
\begin{align}
  S_G = \int d^4x \sqrt{-g} 
  \left[
    f^2 e^{2G}(\partial G)^2 
    + \frac{f^3}{\Lambda^3}(\partial G)^2 \nabla^2 G
    + \frac{f^3}{2\Lambda^3}(\partial G)^4
  \right]~,
\end{align}
The Galilean-gravitation system has a solution describing an emergent universe:
\begin{align}\label{eq:emergent-sol}
  G \simeq -\log(-H_0 t)-\frac{f^2}{2M_p^2 H_0^2 t^2}~,
  \quad
  \rho_G \simeq \frac{f^4}{3M_p^2H_0^4t^6}~,
  \quad
  p_G \simeq -\frac{2f^2}{H_0^2t^4}~,
  \quad
  H \simeq -\frac{1}{3}\frac{f^2}{M_p^2H_0^2t^3}~,   
\end{align}
where $H_0 \equiv \sqrt{2\Lambda^3 / (3f)}$ is a constant. This solution is valid 
at early times
\begin{align}\label{eq:early-approx}
  t^2 \gg \frac{f^2}{M_p^2 H_0^2}~.
\end{align}
Here we have solved the equation of motion of $G$ to second order, and other 
equations to first order. As one can check, this suffices for the calculation of the 
equation of motion for the isocurvaton.

For simplicity, we assume that the Galilean field decays before condition 
(\ref{eq:early-approx}) breaks down. Technically it should be straightforward to 
generalize our analysis to the $t^2\leq f^2/(M_p^2 H_0^2)$ region, provided that
a well defined UV completion is given and the equations of motion are solved
numerically.

As noticed in \cite{Nicolis}, the $G$ field has a blue spectrum, instead of a 
nearly scale invariant spectrum of perturbations. However, it was also
noticed that if a massless scalar field $\phi$ is introduced which couples to the
Galilean like
\begin{align}\label{eq:coupling1}
  S_\phi = -\frac{1}{2}\int d^4x \sqrt{-g} ~ e^{2G} (\partial\phi)^2~,
\end{align}
then in the Galilean background the $\phi$ field feels an effective metric with a 
conformal factor $e^{G/2}$. Inserting the Galilean solution $e^{2G}\simeq 1/(H_0t)^2$, 
it follows that $\phi$ obeys the same equation of motion like a minimally coupled
scalar field in de Sitter space and thus obtains a scale invariant power spectrum if
the fluctuations start out in their quantum vacuum state. 
As discussed in \cite{Laurence}, the ghost condensate can decay to 
excitations of the $\phi$ matter field via the coupling in (\ref{eq:coupling1}),
thus allowing for a graceful exit from the stage of Galileon domination and
onset of the usual radiation phase of Standard Cosmology. 

However, as noticed in \cite{Laurence}, there is a problem to convert the 
perturbation in $\phi$ to curvature perturbations. This is because $\phi$ has 
no potential or background motion. The curvature perturbation couples to the 
energy density of $\phi$, and is thus proportional to $\delta\phi^2$. As a result, 
the curvature perturbation is neither scale invariant nor Gaussian.
If one tries to apply the curvaton scenario \cite{curvaton} and gives the
$\phi$ field a potential which leads to slow rolling, one encounters the
next obstacle: now $\phi$ has a non-vanishing background energy density
which renders the emergent Galilean phase unstable.

Our main point is that
if $\phi$ is an originally massless direction, whose mass is uplifted by the 
Hubble parameter, then there is no such instability since the mass vanishes
deep in the Galileon phase. Here we consider an example realizing this uplifting.

\subsection{An additional scalar with K\"ahler type coupling}

In this subsection, we introduce a model where the additional scalar field arises 
from the K\"ahler potential, or a phenomenological analog. We consider
the action
\begin{align}
  S_{eG} = \int d^4x \sqrt{-g} \left(1 + \frac{\phi^2}{M^2}\right)
  \left\{
    e^{2G} \left[f^2(\partial G)^2 -\frac{1}{2}(\partial\phi)^2\right]
    + \frac{f^3}{\Lambda^3}(\partial G)^2 \nabla^2 G
    + \frac{f^3}{2\Lambda^3}(\partial G)^4
  \right\}~,
\end{align}
where $M$ is a constant with the dimension of mass (could be real or purely 
imaginary). The gravitational sector is standard.

To simplify the calculation, we assume
\begin{align}\label{eq:Gapprox}
  t^2 \gg \frac{f^2}{M_p^2H_0^2}~,\qquad \left|\frac{\phi^2}{M^2}\right|\ll 1~.
\end{align}
The first inequality is discussed in Section \ref{sec:galil-genes-example} and is 
an early time approximation. The second inequality is a condition for the
effective field theory expansion of the new physics to be valid, and it
also ensures that $\phi$ remains 
a sub-dominant component, i.e. a spectator scalar field. 

At the homogeneous and isotropic background level, the matter action for 
$\phi$ takes the form
\begin{align}
  S_{e} = \int d^4 x \frac{1}{H_0^2t^2}
  \left[ 
    \frac{1}{2}\dot\phi^2 - \frac{f^4}{M_p^2 M^2 H_0^2 t^4} \phi^2
  \right]~.
\end{align}
To leading order in the approximations (\ref{eq:Gapprox}), we obtain the solution
\begin{align}
  \phi(t) \simeq \phi_0 - \frac{f^4\phi_0}{5H_0^2 M^2 M_p^2 t^2}~,
\end{align}
where $\phi_0$ is a constant. The second term is much smaller than the first one in a
vast parameter space, considering that $t^2$ is huge at early times. Thus the 
motion of $\phi$ is very slow. 

In the following calculation, we self-consistently treat $\phi$ as a constant. We shall 
come back to the time variation of $\phi$ when discussing the spectral index.

\subsection{Cosmic perturbations}

We start from the ADM metric \footnote{We have chosen a gauge in which the  
spatial metric is flat up to a conformal factor}:
\begin{align}
  ds^2 = -N^2 dt^2 + a^2 e^{2\psi}\delta_{ij}(N^i dt + dx^i)(N^j dt + dx^j)~,
\end{align}
with perturbations $\alpha$ and $\beta$ defined as
\begin{align}
  N = 1+\alpha~, \qquad N_i = \partial_i\beta~.
\end{align}
The matter fields $\phi$ and $G$ are perturbed as
\begin{align}
  \phi \rightarrow \phi + \delta\phi~, \qquad G\rightarrow G + \delta G~.
\end{align}
We have the freedom to set another gauge condition. However for the moment 
we shall not do that and leave the time shift $t\rightarrow t+\delta t$ as a residual 
gauge symmetry.

With the assumptions in equation (\ref{eq:Gapprox}) \footnote{This is to say, neglecting 
terms which are small comparing to other existing terms in the two expansion
parameters.}, one can solve for $\alpha$ and $\beta$ using the constraint equations
and reinsert the solutions into the action.  The second order Lagrangian then
takes the form
\begin{align}
  \mathcal{L}_2 = &
    \left(\frac{3H_0M_p^2t}{f}\right)^2 \left(\dot\psi^2-k^2\psi^2\right)
  \nonumber\\&
  + \left(\frac{f}{H_0t}\right)^2 
    \left(\dot{\delta G}^2-k^2\delta G-\frac{2}{t^2}\delta G^2\right)
  \nonumber\\&
  + \left(\frac{1}{H_0t}\right)^2 
    \left(\frac{1}{2}\dot{\delta\phi}^2-\frac{k^2}{2}\delta\phi^2\right)
  \nonumber\\&
  + \left(\frac{3H_0M_p^2t}{f}\right) \left(\frac{f}{H_0t}\right)
    \left(
      -2\dot\psi\dot{\delta G}
      +\frac{4\delta G\dot\psi}{t}
      +2k^2\psi\delta G
    \right)
  \nonumber\\&
  + \left(\frac{3H_0M_p^2t}{f}\right)\left(\frac{1}{H_0t}\right)
    \left(\frac{4f\phi_0}{3M^2}\right)
    \left(-\dot\psi\dot{\delta\phi}+3\delta\phi\dot\psi+k^2\psi\delta\phi\right)
  \nonumber\\&
  + \left(\frac{f}{H_0t}\right)\left(\frac{1}{H_0t}\right)
    \left(\frac{4f\phi_0}{3M^2}\right)
    \left(
      \dot{\delta\phi}\dot{\delta G} - \delta\phi\dot{\delta G}
      -k^2\delta\phi\delta G
    \right)
  \nonumber\\&
  -\left(\frac{3H_0M_p^2t}{f}\right)^2 
    \left(\frac{3f^4}{2M_p^4H_0^4t^6}\right)\psi^2
  - \left(\frac{3H_0M_p^2t}{f}\right) \left(\frac{f}{H_0t}\right)
    \left(\frac{2f^2}{H_0^2M_p^2t^4}\right)\psi\delta G
~,
\end{align}
Here, the first three lines represent the free Lagrangian, and the 4th to 6th lines are the 
mixing terms. The coefficients of the two terms in the last line are small. We can 
self-consistently neglect these two terms. The self-consistency check involves
obtaining the solution making the approximation, inserting back into the action, 
and checking that the contributions of these two terms are indeed small.

One can define canonically normalized fields via
\begin{align} \label{canonical}
  \sigma \equiv \frac{3H_0M_p^2t}{f} \psi - \frac{f}{H_0t}\delta G ~,
  \qquad \chi \equiv \frac{1}{H_0 t}\delta\phi~.
\end{align}
In terms of these fields the Lagrangian becomes
\begin{align}\label{eq:L-sigma-chi}
  \mathcal{L}_2 = \left( \dot\sigma^2 - k^2\sigma^2 \right)
  + \left( \frac{1}{2}\dot\chi^2 -\frac{k^2}{2}\chi^2 + \frac{\chi^2}{t^2} \right)
  -\frac{4f\phi_0}{3M^2}
  \left(
    \frac{\dot\chi}{t}-\frac{2}{t^2}\chi 
  \right)\sigma~.
\end{align}
Note that $\psi$ or $\delta G$ appear in the above action only in terms of the
gauge-invariant combination $\sigma$. On the other hand, $\chi$ appears in a 
gauge-dependent form. This gauge-dependence is due to the approximation 
$|\phi_0^2/M^2|\ll 1$ which we have made. By this approximation, we are 
asserting that $\phi_0$ is sub-dominant, thus couples to gravity only weakly and 
is treated as a spectator scalar field on a fixed background. At higher
orders, we would recover an explicitly gauge-invariant form. Note that if we
were to choose a uniform $\phi$ gauge, then the parameters in the gauge 
transformation between the gauge we are using and the uniform $\phi$ gauge
will be large, which mixes different orders of the $\phi_0^2/M^2$ terms in the 
second order action.

\begin{figure}[htbp]
\centering
\includegraphics[width=0.7\textwidth]{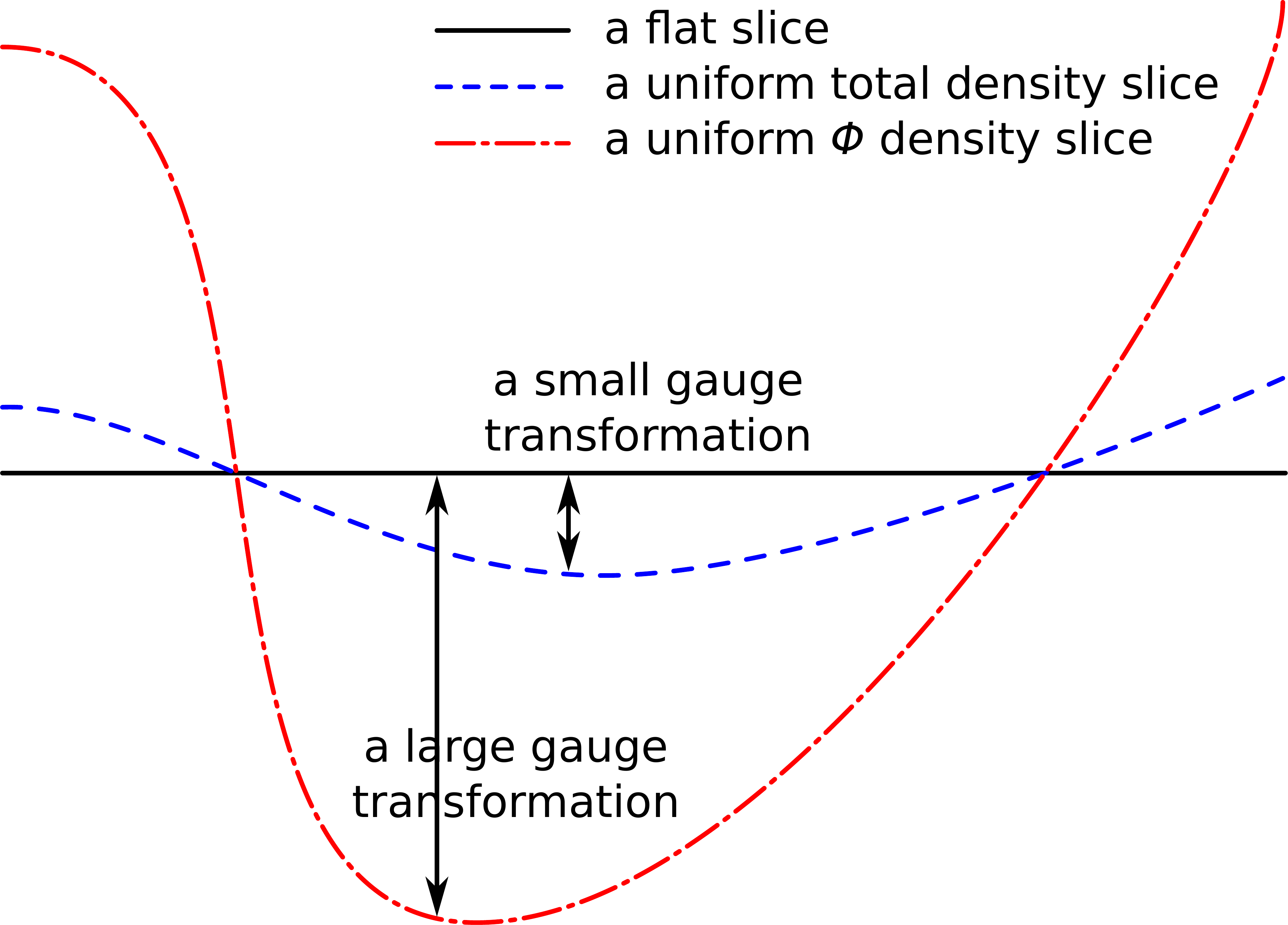}
\caption{Different slices, and ``small'' and ``large'' gauge transformations between them.\label{fig:slices}}
\end{figure}

To put it in another way, the $|\phi_0^2/M^2|\ll 1$ approximation has separated
gauge transformations into two classes, the ``small gauge transformations'' and 
``large gauge transformations'', as illustrated in Figure \ref{fig:slices}. Small 
gauge transformations transform between uniform total energy density slice and the
flat slice, or any other slices not too far away from it. On the other hand, large 
gauge transformations transform between the above slices and the uniform $\phi$ 
gauge. The action (\ref{eq:L-sigma-chi}) can be used when only small gauge 
transformations are considered. While to consider a large gauge transformation, 
one need to use the full second order action, without the approximation 
$|\phi_0^2/M^2|\ll 1$.

Now we make the further assumption that
\begin{align}
  \left|\frac{4f\phi_0}{3M^2}\right| \ll 1
\end{align}
such that the last term in equation (\ref{eq:L-sigma-chi}) can be treated as a perturbation 
when solving the equation of motion of $\chi$. Then the $\sigma$ field has a free 
equation of motion
\begin{align}
  \ddot\chi + k^2\chi - \frac{2}{t^2}\chi & = 0~ .
\end{align}
This is the same equation of motion as a free scalar field $\chi$ in a de-Sitter
background (with time $t$ being replaced by the conformal time). This
demonstrates that $\chi$ will acquire a scale-invariant spectrum in this
background \cite{Nicolis} if it starts out in its vacuum state.

The back-reaction of $\chi$ on $\sigma$ is given by
\begin{align}
  \ddot\sigma + k^2\sigma 
  + \frac{2f\phi_0}{3M^2}\left(\frac{\dot\chi}{t}-\frac{2\chi}{t^2}\right) & = 0~.
\end{align}
In the context of the emergent paradigm it is natural to assume that 
both canonical fields start in their vacuum state. This is similar to the
assumption that scalar matter fields in the de Sitter phase of inflationary
cosmology begin in their vacuum state - although the justification of
these assumptions are very different. With these initial conditions,
the free evolution for $\chi$ leads to the following form of the mode
functions:
\begin{align}
  \chi = \frac{e^{-ikt}}{\sqrt{2k^3}}\left(\frac{1}{t}+ik\right)~.
\end{align}
Inserting this solution into the $\sigma$ equation, we have
\begin{align} \label{sigmaeq}
  \sigma = \frac{e^{-i(kt+\theta_k)}}{\sqrt{k}} + \frac{e^{-ikt}f\phi_0}{\sqrt{2k^3}M^2 t}
  +\frac{ie^{ikt}f\phi_0}{3\sqrt{2k}M^2} \mathrm{Ei}(-2ikt)
  -\frac{ie^{-ikt}f\phi_0}{3\sqrt{2k}M^2}\log(kt)~,
\end{align}
where $\mathrm{Ei}$ stands for the exponential integral function.
Here, the first term comes from the quantum fluctuation of $\sigma$, and the 
rest of the terms are sourced by $\chi$ fluctuations. 
And $\theta_k$ is a relative phase between the original quantum perturbations of the Galilean
and the induced perturbations. Note that the last two 
terms are subdominant because on super-Hubble scales $-kt\ll 1$, 
and thus they are small compared to the second term. For the same reason, 
the second term will dominate over the first term in most of the parameter space.

As discussed above, we can safely work in the uniform total energy density slice, 
in which the curvature perturbation is $\zeta = \psi$. Making use of
(\ref{canonical}) and (\ref{sigmaeq}) and inserting into the definition of
the power spectrum we find the scale-invariant result 
\begin{align}
  P_\zeta(k) = \frac{f^4\phi^2}{36\pi^2 H_\mathrm{eff}^2 M_p^4 M^4 t^4}~,
\end{align}
where $H_\mathrm{eff} \simeq H_0 + f^2/(2M_p^2 H_0 t^2)$ is the effective Hubble 
parameter at Hubble crossing, which can be inferred from the second order 
solution of the equation of motion of $G$ \footnote{From the second 
order action, we can only determine the leading order result with $H_\mathrm{eff} \simeq H_0$. 
However, we can solve the relation between $H_\mathrm{eff}$ and $H_0$ to higher 
orders using the background field equation 
$G \simeq -\log(-H_0 t)-\frac{f^2}{2M_p^2 H_0^2 t^2}$. This second order solution 
is needed if we want to calculate the spectral index. Similarly, we replaced 
$\phi_0$ to $\phi$ in the power spectrum to take into account the slow but non-vanishing 
$k$-dependence.}. This result is true for all times $t$ before the time $t_D$ when
either the Galileon or the field $\phi$ decays. For simplicity, we shall assume that both
fields decay into normal radiation at the same time $t_D$ (e.g. via the process 
studied in \cite{Laurence}). Then, the time $t_D$ is the time when the conversion 
of isocurvature perturbation to curvature fluctuation shuts off. Note that it is the
non-trivial coupling between $\phi$ and $\chi$ introduced in our construction which
leads to the scale-invariance of the curvature fluctuations. The intrinsic spectrum of
curvature fluctuations (in the absence of a $\phi$ field) would be deep blue 
(scalar spectral index $n_s = 3$), as can
be verified by doing the calculation with the first term in (\ref{sigmaeq}) only.

The power spectrum is growing until the $G$ and $\phi$ fields decay. To relate to 
observations, the time appearing in the power spectrum should be taken to be the 
time $t_D$ when the Galileon and spectator scalar field have decayed into
radiation. After the time $t_D$, the isocurvature mode disappears and
the curvature fluctuation remains constant in time (for modes of interest
to us which are far outside the Hubble radius in the radiation phase which
starts after the time $t_D$. 

Note that the increase of the power spectrum in time does not spoil the 
scale invariance because the scale invariance is determined by the $k$-dependence
of the power spectrum. Instead, the increase in time indicates that the perturbation 
in the isocurvature direction is continuously converting to curvature perturbation 
before it decays. This time dependence is similar to what occurs in the curvaton
scenario in inflationary cosmology where the presence of isocurvature fluctuations
see a continuously growing curvature mode \footnote{In the curvaton scenario, where the curvaton $\phi_c$ and the inflaton are decoupled (except via gravity), there is a conserved quantity $\zeta_c = -\psi - H \delta\rho_c / \dot\rho_c$, where $\rho_c$ is the energy density of the curvaton field. The total curvature fluctuation $\zeta$ is not conserved before the decay of curvaton. Instead, it is related to $\zeta_c$ by $\zeta=\zeta_c \dot\rho_c/\dot\rho = \zeta_c (\rho_c+p_c) / (\rho+p)$, where $\rho$ and $p$ are the total energy and total pressure. In our case, $\rho+p\propto t^{-4}$ and $\rho_\phi+p_\phi \propto t^{-6}$ (which is the analog of $\rho_c+p_c$). Thus $\zeta$ scales as $t^{-2}$ and $P_\zeta$ scales as $t^{-4}$. However, there is explicit coupling between $\phi$ and the Galilean. Thus while we hope this footnote still provides an intuitive understanding, an explicit calculation is also needed, as we give in the main text.}.

The amplitude of the resulting spectrum of curvature fluctuations depends on
the values of the mass scales in the Lagrangian, and on the values of $\phi_0$
and $t_D$. It is natural to assume $f \sim M \sim M_{p}$, and to choose 
$\Lambda$ to be parametrically larger than $f$, i.e. $\Lambda = \alpha f$ with
the constant $\alpha \gg 1$. For consistency of the approximations, we must
choose $\phi_0$ parametrically smaller than $M$, i.e. $\phi_0 = \beta M$
with constant $\beta \ll 1$. In this case, we find
\be \label{cond1}
P_{\zeta} \, \sim \, \beta^2 \alpha^{-3} \bigl( M_p t_D \bigr)^{-4}
\ee
which is parametrically smaller than $1$ if $|t_D M_p| \sim 1$.

\begin{figure}[htbp]
\centering
\includegraphics[width=0.45\textwidth]{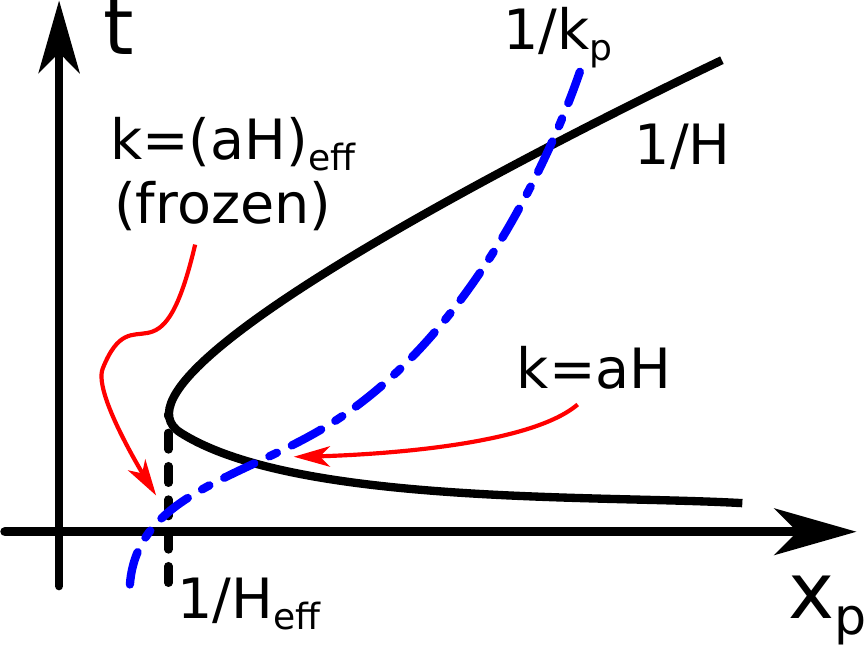}
\caption{The effective Hubble crossing time should be calculated 
as $k=a_\mathrm{eff} H_\mathrm{eff}$ instead of $k=aH$ because the former is 
the time when the fluctuations freeze out, while the latter plays no  special 
role in the equation of motion for the fluctuations.\label{fig:crossing}}
\end{figure}

To calculate the spectral index, we need to compare different $k$ modes. 
As usual, the difference in $k$ is translated into a difference in the time $t(k)$
when the mode $k$ crosses the effective Hubble radius in the emergent
phase. The reason why we must use the effective Hubble radius crossing
$k=a_\mathrm{eff}(t_k) H_\mathrm{eff}(t_k)$
and not the crossing of the background Hubble radius $k = a H$ is that
it is the effective Hubble radius which determines the transition from the
phase when fluctuations oscillate and when they are frozen out
(see Figure \ref{fig:crossing} for an illustration of the scales involved).

The spectral index can be calculated as
\begin{align}
  n_s - 1 = \frac{d\ln P_\zeta}{d\ln k} = -t \frac{d\ln P_\zeta}{dt}
  = -2t \frac{\dot\phi}{\phi} + 2t \frac{\dot H_\mathrm{eff}}{H_\mathrm{eff}}~,
\end{align}
Inserting the equations of motion, we have
\begin{align}\label{eq:index}
  n_s-1 = - \left(\frac{2f^2}{5M^2}+1\right) \frac{2f^2}{M_p^2H_0^2t_k^2} ~.
\end{align}
The spectral index could be extremely small. This is because we want to 
generate a large hierarchy of scales for which the spectrum is scale invariant.
Let the number of e-foldings of the effective de Sitter phase be $N$. Then 
we need
\begin{align}
  \frac{t_k}{t_D} = 
  \frac{a_\mathrm{eff}(t_D)H_\mathrm{eff}(t_D)}{a_\mathrm{eff}(t_k)H_\mathrm{eff}(t_k)}
  = e^N.
\end{align}
Thus, the spectral index is very small given a large $N$. This can be verified by inserting the above expression into (\ref{eq:index}). We get
\begin{align} \label{cond2}
  \alpha^{3/2} M_p t_D \simeq \frac{1}{e^N\sqrt{1-n_s}} ~.
\end{align}
Present experiments suggest that $1-n_s$ is a few percent. 
In principle, the free parameters $\alpha$, $\beta$ and $t_D$ can be
chosen such that the power spectrum amplitude (\ref{cond1}) agrees
with observations and the tilt reproduces the best-fit value (\ref{cond2}). 
However, it may be difficult to fit the current data of $n_s$ 
because the presence of the exponential factor $e^{N}$. 
Additional mechanisms may be needed to generate a tilt of the power spectrum.

\section{Pseudo-Conformal Model}

Similarly, for the pseudo-conformal model of \cite{Khoury}, we consider the 
matter action
\begin{align}
  S_m = \int d^4 x \sqrt{-g} \left( 1+\frac{\phi^2}{M_1^2}\right) 
  \left[ 
    -\frac{1}{2} (\partial C)^2 + \frac{\lambda}{4}C^4 
    -\frac{C^2}{2M_2^2} (\partial\phi)^2
  \right]~,
\end{align}
where $C$ is the pseudo-conformal field, which drives the background 
motion of the emergent cosmology. The factor $(1+\phi^2/M_1^2)$ is 
again the new factor which we are adding, a factor which is 
motivated by the K\"ahler potential or its phenomenological analog.

At the background level, the dynamics of the scale factor, and of the 
fields $C$ and $\phi$ goes as
\begin{align}
H=\frac{1}{3\lambda t^3 M_p^2}~, \qquad  C = \sqrt{\frac{2}{\lambda}} ~\frac{1}{t}~, \qquad \phi = \phi_0 t^{-\frac{2M_2^2}{3M_1^2}}~.
\end{align}
One can set up cosmic perturbations following what was done in  the 
previous section. Self-consistently, we postulate that the following quantities 
are small
\begin{align}
  1/(6\lambda t^2 M_p^2)~, \qquad |\phi^2/M_1^2|~, \qquad
  \left|M_2\phi_0 / M_1^2\right|~, \qquad |M_2^2 / M_1^2|~.
\end{align}
The first three quantities are analogs of conditions imposed in the Galilean 
Genesis example, corresponding to early time, sub-domination of $\phi$, and 
weak coupling of fluctuations between the two degrees of freedom, respectively. 
The requirement of the smallness of the 4th quantity is new. It comes about
as follows: If $|M_2^2/M_1^2|\geq 1$, then the $\phi$ field will roll down its 
emergent potential too quickly, and no scale invariant spectrum will be generated.

With the above assumptions, the ``small'' gauge invariant quantities for
fluctuations can be constructed as
\begin{align}
  \sigma \equiv 3M_p^2t\sqrt{\lambda} \psi - \frac{\delta C}{\sqrt{2}}~,
  \qquad \chi \equiv \frac{\sqrt{2}\delta\phi}{M_2 t \sqrt{\lambda}}~.
\end{align}
Using these definitions, the second order Lagrangian can be written as
\begin{align}\label{eq:L-sigma-chi-conf}
  \mathcal{L}_2 = \left( \dot\sigma^2 - k^2\sigma^2 \right)
  + \left( \frac{1}{2}\dot\chi^2 -\frac{k^2}{2}\chi^2 + \frac{\chi^2}{t^2} \right)
  +\frac{2\sqrt{2}M_2\phi_0}{3M_1^2}
  \left(
    \frac{\dot\chi}{t}-\frac{2}{t^2}\chi
  \right)\sigma~.
\end{align}
Interestingly, the pseudo-conformal case and the Galilean case share 
the same structure of the perturbations, except for a different sign in the 
interaction term, which is only a different convention and thus unimportant 
\footnote{Technically, there are actually two differences between the calculations 
in the Galilean case and in the pseudo-conformal case. First, in the pseudo-conformal 
example, although the time evolution of $\phi$ is still slow, it is now of the same 
order as the interaction term in the second order Lagrangian (the third term in 
Equation (\ref{eq:L-sigma-chi-conf})). Thus the time evolution of $\phi$ cannot 
be neglected. In the Galilean case, $\phi$ can be treated as a constant. 
Second, in deriving the interacting term in Equation (\ref{eq:L-sigma-chi-conf}), 
we have used the free field equation of motion for $\chi$. Strictly speaking, we are 
not allowed to use the equation of motion in deriving the action. However, here the 
use of the equation of motion should be understood as a redefinition of the 
perturbation variables, as in the case of the field redefinition done in 
\cite{Maldacena:2002vr}.}.

Thus we have the equations of motion in the Born approximation:
\begin{align}
  \ddot\chi + k^2\chi - \frac{2}{t^2}\chi & = 0~,
\end{align}
\begin{align}
  \ddot\sigma + k^2\sigma 
  - \frac{2\sqrt{2}M_2\phi_0}{3M_1^2}\left(\frac{\dot\chi}{t}-\frac{2\chi}{t^2}\right) & = 0~.
\end{align}
Most naturally, the fields should start from a vacuum state. Thus
\begin{align}
  \chi = \frac{e^{-ikt}}{\sqrt{2k^3}}\left(\frac{1}{t}+ik\right)~.
\end{align}
Inserting the solution into the $\sigma$ equation, we have
\begin{align}
  \sigma = \frac{e^{-i(kt+\theta_k)}}{\sqrt{k}} - \frac{e^{-ikt}M_2\phi_0}{\sqrt{k^3}M_1^2 t}
  -\frac{ie^{ikt}M_2\phi_0}{3\sqrt{k}M_1^2}\mathrm{Ei}(-2ikt)
  +\frac{ie^{-ikt}M_2\phi_0}{3\sqrt{k}M_1^2}\log(kt)~,
\end{align}
where $\theta_k$ is a relative phase between the original quantum perturbations of $C$
and the induced perturbations.

As in the previous section, the second term dominates the $-kt\ll 1$ regime. 
Thus, the power spectrum is
\begin{align}
  P_\zeta = \frac{M_2^2 \phi_0^2}{18\pi^2 \lambda M_p^4 M_1^4 t_D^4}~,
\end{align}
where $t_D$ is the time when the pseudo-conformal phase ends and
the isocurvature field decays.
Again, the spectral index is small and the spectrum is nearly scale invariant.

\section{Conclusions}

We have shown how correction terms to the effective action of the low
energy fields which are expected based on supergravity or quantum
gravity considerations induce a linear coupling between the fluctuations
of an isocurvature field and the curvature fluctuations. Thus, a
scale-invariant spectrum in a spectator scalar field induces a
scale-invariant curvature fluctuation spectrum. Scale-invariant
spectra for spectator scalar fields are induced e.g. in
the Galileon Genesis model \cite{Nicolis}, in the Conformal Cosmology
of \cite{Rubakov} and in the Pseudo-Conformal scenario of
\cite{Khoury}. Our mechanism can be used to demonstrate that
these scenarios indeed lead to a scale-invariant spectrum of
curvature perturbations.

\begin{acknowledgments}

We are grateful to Yi-Fu Cai and Laurence Perreault-Levasseur for stimulating
discussions. This research was supported in part by an NSERC Discovery grant 
and by funds from the Canada Research Chair program.

\end{acknowledgments}

\end{document}